\documentclass[namedreferences]{solarphysics}

\usepackage[hyperref,optionalrh]{spr-sola-addons} 
\usepackage{graphicx}        
\usepackage{color}           

\chardef\us=`\_

\begin{document}

\begin{article}

\begin{opening}

\title{Deriving Large Coronal Magnetic Loop Parameters Using LOFAR J burst Observations}



\author{Jinge ~\surname{Zhang}$^{1}$\sep
        Hamish A. S. ~\surname{Reid}$^{1}$\sep
        Vratislav ~\surname{Krupar}$^{2,3}$\sep
        Pietro ~\surname{Zucca}$^{4}$\sep
        Bartosz ~\surname{Dabrowski}$^{5}$\sep
        Andrzej ~\surname{Krankowski}$^{5}$
      }
\runningauthor{J. Zhang et al.}
\runningtitle{Deriving large coronal magnetic loop parameters using LOFAR J burst observations}

  \institute{$^{1}$ Mullard Space Science Laboratory,   University College London, RH5 6NT, United Kingdom
                     email: \url{jinge.zhang.18@ucl.ac.uk} 
                     email: \url{hamish.reid@ucl.ac.uk}\\
              $^{2}$ Goddard Planetary Heliophysics Institute, University of Maryland, Baltimore County, Baltimore, MD, USA
                     email: \url{vratisla@umbc.edu} \\
              $^{3}$ Heliophysics Science Division, NASA Goddard Space Flight Center, Greenbelt, MD, USA
                     email: \url{} \\
              $^{4}$ ASTRON, The Netherlands Institute for Radio Astronomy, Oude Hoogeveensedijk 4, 7991 PD Dwingeloo, The Netherlands
                     email: \url{zucca@astron.nl} \\
              $^{5}$ Space Radio-Diagnostics Research Centre, University of Warmia and Mazury, R. Prawochenskiego 9, 10-719 Olsztyn, Poland
                     email:\url{bartosz.dabrowski@uwm.edu.pl}
                     email:\url{kand@uwm.edu.pl}\\
              }

\begin{abstract}
Large coronal loops around one solar radius in altitude are an important connection between the solar wind and the low solar corona.  However, their plasma properties are ill-defined as standard X-ray and UV techniques are not suited to these low-density environments.  Diagnostics from type J solar radio bursts at frequencies above 10\,MHz are ideally suited to understand these coronal loops.  Despite this, J bursts are less frequently studied than their type-III cousins, in part because the curvature of the coronal loop makes them unsuited for using standard coronal density models.  We used $LOw-Frequency-ARray$ (LOFAR) and $Parker\ Solar\ Probe$ (PSP) solar radio dynamic spectrum to identify 27 type-III bursts and 27 J bursts during a solar radio noise storm observed on 10 April 2019.  We found that their exciter velocities were similar, implying a common acceleration region that injects electrons along open and closed magnetic structures.   We describe a novel technique to estimate the density model in coronal loops from J burst dynamic spectra, finding typical loop apex altitudes around 1.3 solar radius.  At this altitude, the average scale heights were 0.36 solar radius, the average temperature was around 1 MK, the average pressure was 0.7\,mdyn cm$^{-2}$, and the average minimum magnetic field strength was 0.13\,G.  We discuss how these parameters compare with much smaller coronal loops.  
\end{abstract}
\end{opening}


\section{Introduction}
     \label{sec:introduction} 

During the early stages of researching solar radio bursts, \citet{maxwell1958new} observed a new spectral characteristic in solar radio emissions with an inverted ``U" shape on dynamic spectra. Since then, this type of burst has been called the type U solar radio burst. If there is no emission on the second half of the burst (descending leg), such bursts are called inverted type J or simply J bursts. J and U bursts are generally known as fast-drift solar radio bursts \citep[e.g.][]{kruger2012introduction} according to their rapid frequency drift from high to low frequency, although U bursts also have a positive drift rate on the descending leg. Both J and U bursts are generally believed to be generated by accelerated electron beams propagating along closed magnetic loops in the solar corona \citep[e.g.][]{klein1993observation,karlicky1996transport}. Similar to J bursts, the most common fast-drift burst is the type-III solar radio burst \citep[see review by][]{reid2014review}, which is generally accepted to be produced by electron beams traveling along ``open" flux tubes.

The plasma emission mechanism based on the work of \citet{ginzburg1959mechanisms} is the most accepted radio emission process for the majority of these coherent fast-drift solar radio bursts.  Electron beams generate Langmuir waves near the local background plasma frequency.   These Langmuir waves can subsequently coalesce with ion sound waves and are converted into electromagnetic waves at the same frequency, a process known as fundamental emission.  Langmuir waves can also coalesce with other Langmuir waves travelling nearly in the opposite direction to produce electromagnetic waves at double the Langmuir wave frequency, a process known as harmonic emission. One signature of harmonic emission is the low degree of circular polarization, around 10\,\% \citep{dulk1980position}, with fundmental emission having a higher average circular polarization, around 35\,\%.  During solar type-III burst storms, fast-drift bursts are usually observed with a low degree of polarization, indicating that harmonic emission is more common \citep[e.g.][]{kai1985storms}. The degree of circular polarization of U bursts between 10\,--\,300\,MHz is generally low, which suggests that J and U bursts are more commonly produced by harmonic emission in this frequency range \citep[e.g.][]{labrum1970solar,aurass1997spectrographic}.

Thanks to the development of the plasma emission mechanism theory, fast-drift radio bursts are important diagnostic tools for probing solar coronal loop characteristics when the plasma density is too low for standard EUV and X-ray analysis, around $10^{7.5}~\rm{cm}^{-3}$ at altitudes around one solar radius and above.  For the low background plasma densities at these altitudes, we note that Coulomb collisions do not play a significant role in electron beam kinetics.  As well as coronal loop characteristics, we can also use these radio bursts to identify the properties of the electron beams that drive radio burst emission.

The electron beam velocity can be estimated by knowing the exciter travel distances by estimating a coronal density model, and the time duration from the dynamic spectra.  In general, type-III bursts exciter velocities are generally accepted between 0.1 and 0.5\,c \citep[e.g.][]{reid2014review,carley2016radio}. \citet{poquerusse1994relativistic} estimated exciter velocities of normal type-III bursts (in range 100 to 500\,MHz) of ${v_\mathrm{e} \approx 0.3\,\rm{c}}$, where c is the speed of light, by assuming a density model with scale height ${H \lesssim 10^5\,\rm{km}}$. They found velocities of electron beams generate type-IIId bursts close to the speed of light because they have shorter observed characteristic frequency drift time (${\approx 0.7\,\rm{second}}$). Similarly, type-III bursts exciter speed estimated by \citet{dulk1987speeds} in the frequency between 30 and 1980\ kHz had an average speed of 0.14\,c, which is close to \citet{klassen2003superluminal}'s estimation that the average exciter speed of interplanetary (IP) type-III bursts is 0.15\,c. Between frequencies 30 to 70\,MHz, \citet{reid2018solar} selected 31 type-III bursts observed by $LOw-Frequency-ARray$ (LOFAR) and derived average exciter velocities from the peak time of each radio burst is around 0.17\,c in average. 

Estimating exciter velocities cannot be directly applied on J bursts. This is because the curvature part of the J burst dynamic spectrum is due to an accelerated electron beam travelling along the bent part of the closed loops, where standard numerical coronal density models cannot be applied. Many studies determined J burst exciter velocities using different methods. \cite{labrum1970solar} analysed 29 clearly defined U-bursts in the 10\,--\,200\,MHz range and estimated average exciter speed at 0.25\,c by using time duration measured at double of the turnover frequency. \cite{reid2017imaging} measured three J and U bursts' exciter velocities by estimating the exciter positions using radio images, finding average speed determined from the peak time of the radio bursts was 0.21\,c. At higher frequency ranges, other studies estimated the beam velocity between 0.16\,c to 0.53\,c using the Coulomb collisional time \citep[e.g.][]{yao1997solar,wang2001observations,fernandes2012flaring}.

Based upon exciter velocity studies, it is generally considered that J and U's exciter velocities have the same order as exciter velocities from type-III bursts \citep[e.g.][]{labrum1970solar}. However, what is not clear is whether the same active region would produce accelerated electron beams with similar velocities near the footpoint of magnetic loops confined to the corona (emitting J bursts) and magnetic loops that extend into the heliosphere (emitting type-III bursts).  Different properties might be expected in these magnetic loops have very different plasma properties near the footpoint of the loops where the electron acceleration processes occur.

Other than exciter velocities, the physical parameters of coronal loops can be determined by analysing J or U bursts.  For example, \citet{Aschwanden:1992aa} analysed three U bursts detected by the $Very\ Large\ Array$ (VLA) in the range of 1.1 \,--\,1.7\,GHz in August 1989. After measuring average electron beam velocities of 53\,Mm/s (0.18\,c), \citet{Aschwanden:1992aa} derived loop apex altitudes around 130 Mm and determined the upper limit for the density scale height was ${\approx 370\,\rm{Mm}}$. \citet{Aschwanden:1992aa} provided another way of determine the scale height value, using the definition of scale height for the coronal plasma in thermal equilibrium and they derived the scale height from the observed loop apex temperature. Many high frequency U and J bursts studies derived the loop pressure by using the Rosner--Tucker--Vaiana Law \citep[RTV Law:][]{rosner1978dynamics}, which describes the relationship between the loop size, temperature and pressure \citep[e.g.][]{Aschwanden:1992aa,yao1997solar,wang2001observations,fernandes2012flaring}. Moreover, the lower limit of the magnetic field strength is mostly determined by applying the plasma $\beta$ formula, which is the ratio of thermal to magnetic pressure. 

Although \cite{Aschwanden:1992aa} provides a good example for determining coronal loop physical parameters and electron beam properties from U bursts, the coronal loop is low in altitude due to the high observational frequency range provided by VLA. Lower frequency range J bursts have been studied in a similar way as well. Recently, \citet{dorovskyy2021solar} analysed a U burst observed by $Giant\ Ukrainian\ Radio\ telescope$ (GURT) on 18 April 2017 in the frequency band 10\,--\,80\,MHz. \citet{dorovskyy2021solar} assumed a beam speed similar to observed type-III exciter speeds at 0.2\,c for fundamental emission and 0.33\,c for second-harmonic emission.  \citet{dorovskyy2021solar} then estimated the density profile of the closed coronal loop by assuming loop temperature at 1.4\,Mk (and twice as high temperature 2.8\,Mk) from \cite{mann1999heliospheric}'s measurement.

We analyse type-III and J bursts observed during a solar radio noise storm. In Section \ref{sec:Velocities of electron beams}, we analysed 27 type-III and 27 J bursts observed by LOFAR between 20 to 80\,MHz.  We classified radio bursts by combining LOFAR and PSP dynamic spectrum to check continuities of type-III bursts and drift rate of J bursts. Then, we determined exciter velocities from the drift rate of type-III bursts and the high frequency part (type-III-like part) of J bursts to make the comparison. In Section \ref{sec:Physical parameters of coronal loops}, we analyse 24 selected J bursts during the radio noise storm and inferred ambient plasma density model of coronal loops by assuming the exciter velocity remains constant while travel along closed flux tubes and loop top geometry is semi-circular in shape. Then we estimated coronal loop parameters (including temperature, pressure and minimum magnetic field strength) distributions of the loop top by using the density scale height value from the inferred density model. In Section \ref{sec:Discussions}, we discuss the results of the comparison between exciter velocities propagated along the ``open" and ``closed" coronal loops.  We also compare the physical parameters we estimated for large coronal loops to smaller loops estimations studied by \citet{Aschwanden:1992aa}.  Moreover, we discuss factors that affect coronal loop physical parameters estimations in this work. 

\section{Instruments and Observations} \label{sec:Instruments and observations}

\subsection{LOw-Frequency-ARray (LOFAR) Observations}\label{sec:SRBS LOFAR}

We focus on a solar radio burst noise storm, which was observed on the 10th of April 2019 from 13:42 to 15:39 UT by the $LOw-Frequency-ARray$ (LOFAR) \citep[LOFAR:][]{van2013lofar} using the low band antennas (LBA) that operate between 20 and 80\,MHz. We used the data from the LOFAR observation project \textsf{$\rm{LT}10\_002$}.  LOFAR provides high resolution spectral data with a sub-band width of 0.192\,MHz and a time resolution of 0.01\,s. To improve the signal to noise ratio for the radio burst spectroscopy analysis, we integrate the time resolution to approximately 0.1\,s. The 60\,MHz bandwidth was covered irregularly using 60 sub-bands.  Type-III and J bursts were mostly observed during the solar radio burst noise storm, which provides opportunities to statistically analyse both types of radio bursts during the same solar activity event. 

\subsection{Parker Solar Probe (PSP) observations}\label{sec:The Parker Solar Probe observation}

$Parker\ Solar\ Probe$ (PSP) is a NASA mission that travels much closer to the Sun than any previous spacecraft in human history. On the PSP spacecraft, the science instrument FIELDS takes measurements of magnetic fields, plasma waves and turbulence, and radio emissions in the inner heliosphere \citep[see][]{bale2016fields}.  We used data measured by the $Radio\ Frequency\ Spectrometer$ (RFS), which is a spectrometer in the FIELDS. The radio noise storm occurred during the second PSP perihelion observing campaign and so FIELDS was in burst mode, providing the time--frequency spectrum of radio flux observations with a temporal resolution of seven seconds. The high-frequency receiver (HFR) has 64 frequency channels between 1.3 and 19.2\,MHz \citep[see][e.g.]{krupar2020density}.

\begin{figure*}  \center  
\includegraphics[width=0.8\textwidth]{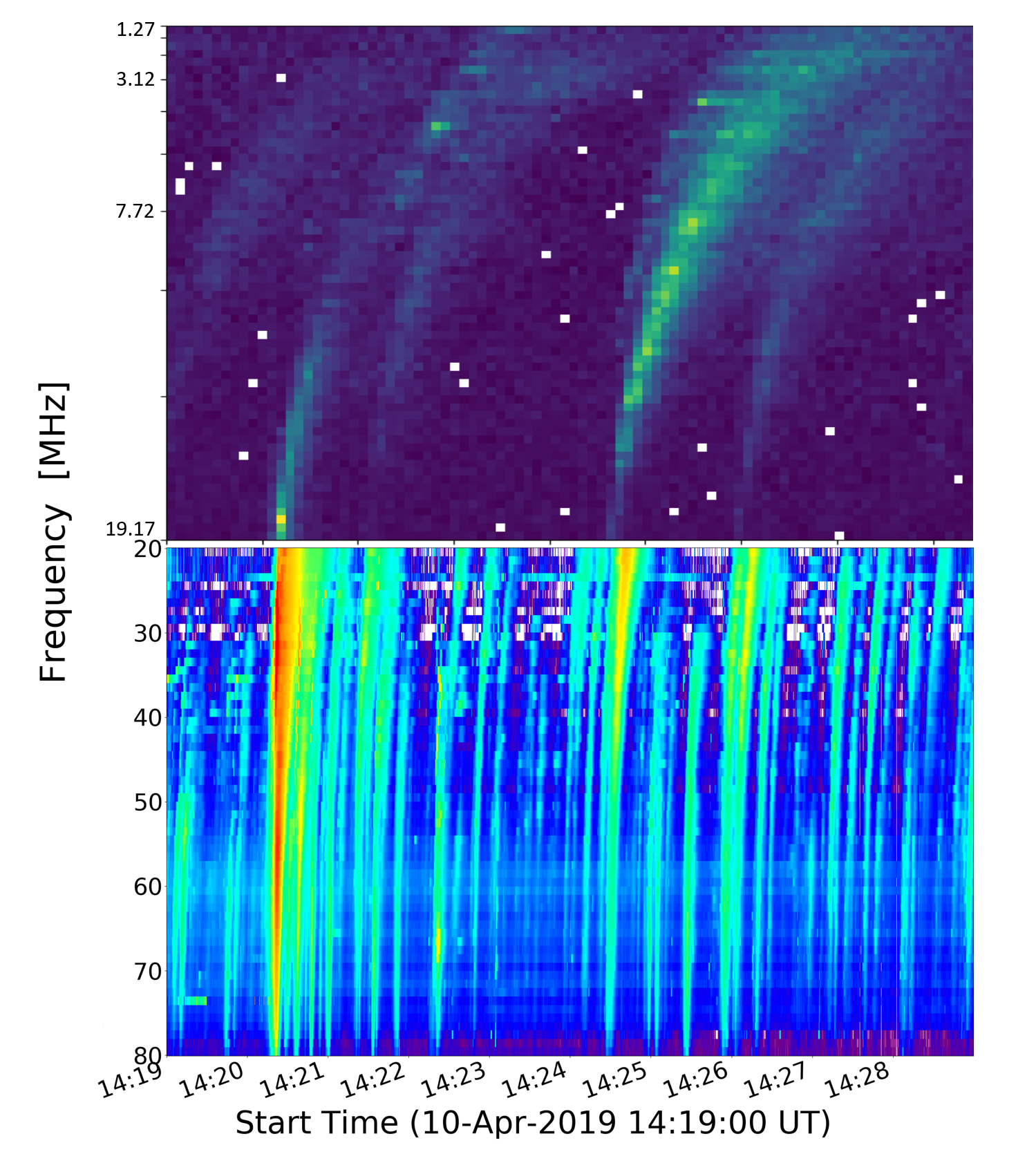}
\caption{Ten minutes of data combining LOFAR and PSP dynamic spectrum, used for classifying type-III and J bursts during the noise storm on the 10th April 2019. $The top panel$: PSP RFS radio observations between 1.3\,--\,19.2\,MHz; $The bottom panel$: LOFAR LBA observation between 20\,--\,80\,MHz. There are five interplanetary (IP) type-III bursts identified in the PSP dynamic spectrum that are also observable by LOFAR.}
\label{fig:pspLOFARcomb}
\end{figure*}

The combination of PSP and LOFAR measurements provide a relatively larger scale picture of the accelerated electron beam transport and the coronal structure along its travel path. We used data from RFS/HFR because its highest frequency channel (19.7\,MHz) is close to LOFAR LBA's lowest channel (20\,MHz), which provides the best continuities between the two spectra. An example of ten minutes of data during the radio noise storm is presented in Figure \ref{fig:pspLOFARcomb}. On 10 April 2019, PSP was 0.25\,AU away from the Sun and 0.78\,AU away from the Earth. The radio signal took approximately six minutes to travel to LOFAR after being observed by PSP.  We corrected the time difference between the PSP and LOFAR measurements when we combined the dynamic spectrum in Figure \ref{fig:pspLOFARcomb}.

\section{Electron Beam Velocities}\label{sec:Velocities of electron beams}

\subsection{Classification of type-III and J Bursts}\label{sec:Classification of type III and type J bursts}

During the noise storm, fast drift solar radio bursts were observed on the LOFAR dynamic spectrum between 20 to 80\,MHz, and the PSP observed interplanetary (IP) type-III solar radio bursts between 1.3 to 19.2\,MHz. We classified a type-III burst as a radio burst that was observed both by LOFAR and by PSP.  For example, from the ten minutes of data in Figure \ref{fig:pspLOFARcomb}, we can see on the PSP dynamic spectrum (upper panel) there are five radio bursts that can be identified as type-III bursts that extend below 20\,MHz.  As Figure \ref{fig:pspLOFARcomb} shows, the lower sensitivity of PSP means that many weaker type-III bursts were observed in LOFAR but not in PSP, and they were ignored for this study.  We classified a J burst as a radio burst that showed a clear turnover frequency in the LOFAR dynamic spectrum before 20\,MHz. By combining the PSP and LOFAR observations in different wavelength ranges, we identified 27 type-III bursts and 27 J bursts. 

\subsection{Velocity Estimations} \label{sec:drift rates and velocities}

Whilst electron beam velocities have been estimated from type-III bursts and J bursts individually in many studies, electron beams velocities have never been simultaneously reported during the same radio noise storm.  Despite having similar reported velocities, we aim to identify here whether there is any systematic trend between electron beams accelerated in closed and open flux tubes.  The electron beam (exciter) velocity is related to the frequency drift rate of radio bursts via

\begin{equation} \label{eqn:Dr and vel}
\frac{\mathrm{d}f_\mathrm{e}}{\mathrm{d}t} = \frac{\mathrm{d}f_\mathrm{e}}{\mathrm{d}n_\mathrm{e}}\frac{\mathrm{d}n_\mathrm{e}}{\mathrm{d}l}\frac{\mathrm{d}l}{\mathrm{d}t},
\end{equation}

\noindent
where $f_\mathrm{e}$ is the plasma frequency, $\frac{\mathrm{d}f_\mathrm{e}}{\mathrm{d}t}$ is the frequency drift rate, $n_\mathrm{e}$ is the background electron density, $l$ is the path of the electron beam, and $\frac{\mathrm{d}l}{\mathrm{d}t}$ is the exciter velocity.

\begin{figure*} \center  
 \includegraphics[width=0.9\textwidth,trim=8 0 20 10,clip]{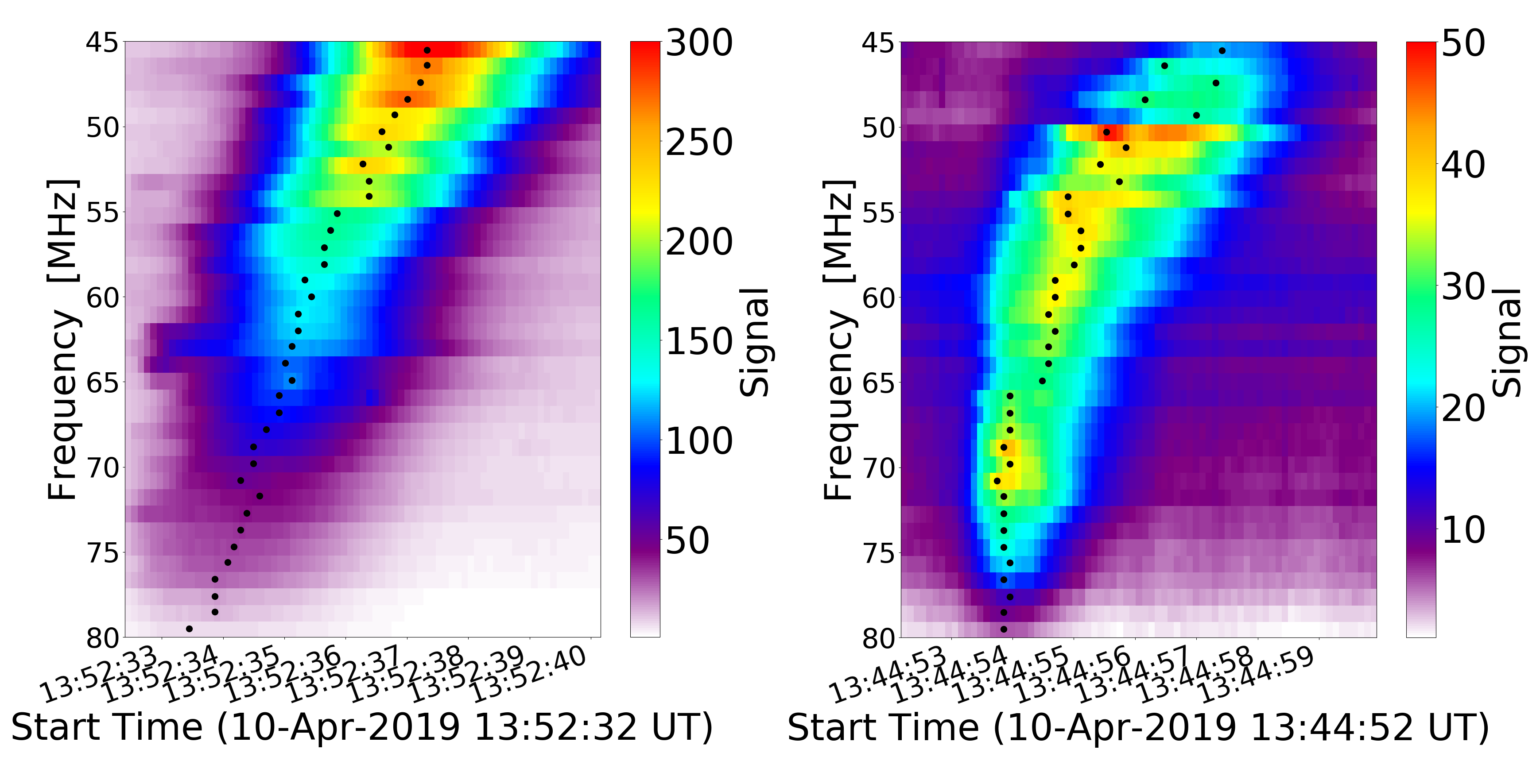}
\caption{type-III burst (\textbf{left}) and a J burst (\textbf{right}) on the dynamic spectrum. type-III and J bursts show rapid drift from high to low frequencies. type-III bursts have relatively constant frequency drift rate, and J bursts show decreased frequency drift rate. \textbf{Black dots} on both dynamic spectra are maximum flux points for each frequency band, and their time profiles show frequency drift.}
\label{fig:TJ and T3 dynamic spectrum} 
\end{figure*}

We identified the maximum flux points for each frequency sub-band for the selected bursts on the LOFAR dynamic spectrum to determine the drift rate.  Figure \ref{fig:TJ and T3 dynamic spectrum} shows an example of a type-III burst (left) and a J burst (right) between 45 to 80\,MHz.  The black dots are the maximum flux points in time.  Note the nearly constant frequency drift rate for the type-III burst between 80 and 45\,MHz, and the change in the frequency drift rate for the J burst around 50\,MHz.   These maximum flux points provide the temporal and frequency profile of the bursts drifting on the dynamic spectrum. By assuming fundamental or second-harmonic emission, we converted the observational frequencies $f_{\mathrm{obs}}$ to the ambient plasma frequencies $f_\mathrm{e}$ ($f_{\mathrm{obs}} = f_\mathrm{e}$ for the fundamental emission, $f_{\mathrm{obs}} = 2 f_\mathrm{e}$ for the second-harmonic emission).  Then we converted the plasma frequency to the ambient plasma density and applied a solar coronal plasma density model to determine associated distances, and hence we obtained the exciter velocities of our identified type-III bursts.  We assumed the  Saito model \citep{saito1977study} times a factor of 4.5. The multiplier of the density model's magnitude was derived by \citet{reid2017imaging}, who derived three empirical density models multiplied by constant factors to fit the three J/U bursts density distributions.  The Saito $\times$ 4.5 density model is given by

\begin{equation} \label{eqn:dens_model}
    n_\mathrm{e}(r) = 4.5 \times \left(1.3\times10^6\left[\frac{r}{\mathrm{R_\odot}}\right]^{-2.14} + 1.68\times10^8 \left[\frac{r}{\mathrm{R_\odot}}\right]^{-6.13}\right),
\end{equation}
where $r$ is the distance from the centre of the Sun, in cm.

The J bursts have exciters that are propagating along the apex of a coronal loop.  Therefore the standard coronal density model, which assumes an open flux tube does not capture how the corresponding plasma density changes at the apex of these closed coronal loops as the geometry is different.  Therefore we defined the ``type-III-like" part of the J bursts as the ascending leg where the drift rate remains roughly constant. For the J burst shown in Figure \ref{fig:TJ and T3 dynamic spectrum} this was above 60\,MHz.  We then found exciter velocities from the J burst maximum flux points in this higher frequency range, where there was no obvious change in the frequency drift rate, using the same density model as we used for the type-III bursts.

Table \ref{tab:velocities} shows the average exciter velocities found by analysing all the identified type-III and J bursts.  Under the assumption of fundamental emission mechanism, type-III bursts average exciter velocity is 0.16\,c which is very close and within the standard deviation of the J bursts' average at 0.17\,c.  If all identified bursts are assumed as second-harmonic emission, type-III burst exciters' average velocity is 0.18\,c which again is within one standard deviation of the J burst average exciter velocity of 0.22\,c.  Although type-III and J bursts' exciter velocities are close, there is a systematic increase in the J burst exciters compared to the type-III burst exciter velocities.

\begin{center}
\begin{table*} \centering
\caption{Average exciter velocities of all identified type-III and J bursts.}
\begin{tabular}{ c c c }

\hline\hline

&	Fundamental emission   &   Second-harmonic emission    	          \\
&	Average velocity [c]  &	Average velocity [c] 	\\ \hline
27 type-III bursts  &	0.16 $\pm$ 0.03 	         &	0.18 $\pm$ 0.05	\\ 
27 J bursts  &	0.17 $\pm$ 0.04 	         &	0.22 $\pm$ 0.05	\\
 \hline
\end{tabular}
\label{tab:velocities}
\end{table*}
\end{center}

\section{Physical Parameters of Coronal Loops} \label{sec:Physical parameters of coronal loops}

\subsection{Coronal Loop Density Model}\label{sec:Inferred background density model for the curvature part of the loop}

The change in the drift rate of the J bursts at the turnover frequencies provides information about the change in electron density at the apex of the closed magnetic loops.  We can therefore use the J burst drift rates to estimate the background electron density of large coronal loops.  

We start by estimating the travel distance of the electron beam along the apex of the coronal loop.  Assuming that the exciter velocity does not change along the loop, we determine the travel distance using the exciter velocity, $v_{\rm exciter}$, calculated from the ``type-III-like" part of the J burst, and the times, $t$ of the maximum flux points in the J burst. The travel distance along the coronal loop $l(t)$ is given by

\begin{equation} \label{eqn:position}
l = l_0 + v_{\rm exciter} (t-t_0),
\end{equation}
where $l_0$ is the reference altitude of the source position, derived from the highest frequency of the J burst and our assumed density model (Saito model $\times$ 4.5). The start time $t_0$ is the time of the max flux point at this frequency, and the term $t-t_0$ is the travel time of exciter along the apex of the magnetic loop. 

The travel distance along the loop, $l$, does not reflect the altitude of the loop apex.  If we want to know the loop altitude, we must assume the loop geometry.  In this study, we assume the loop top is a semi-circle.  We then converted the travel distance profile along the loop $l$ to a solar altitude $h$. 

We determined the coronal loops electron density distributions with solar altitude $n_\mathrm{e}(h)$ using the times of maximum flux at each frequency and the exciter profile in solar altitude as a function of time.  Assuming that the coronal loop is in hydrostatic equilibrium, we then model the $n_\mathrm{e}(h)$ with an exponential of the form 

\begin{equation}  \label{eqn:exponential}
n_\mathrm{e}(h) = n_0 \exp{\left(\frac{-h}{\lambda}\right)},
\end{equation}

\noindent
where $n_0$ is the reference plasma density, a constant derived from density associated with the height $l_0$, and $\lambda$ is the hydrostatic density scale height.

During this solar radio storm, we identified 24 J bursts for inferring the coronal density model and deriving large coronal loops' physical parameters. These J bursts have frequency drift rates that decreases in the lower frequency range and relatively noiseless dynamic spectrum backgrounds.

\begin{figure*} \center  
 \includegraphics[width=0.7\textwidth,trim=8 0 20 10,clip]{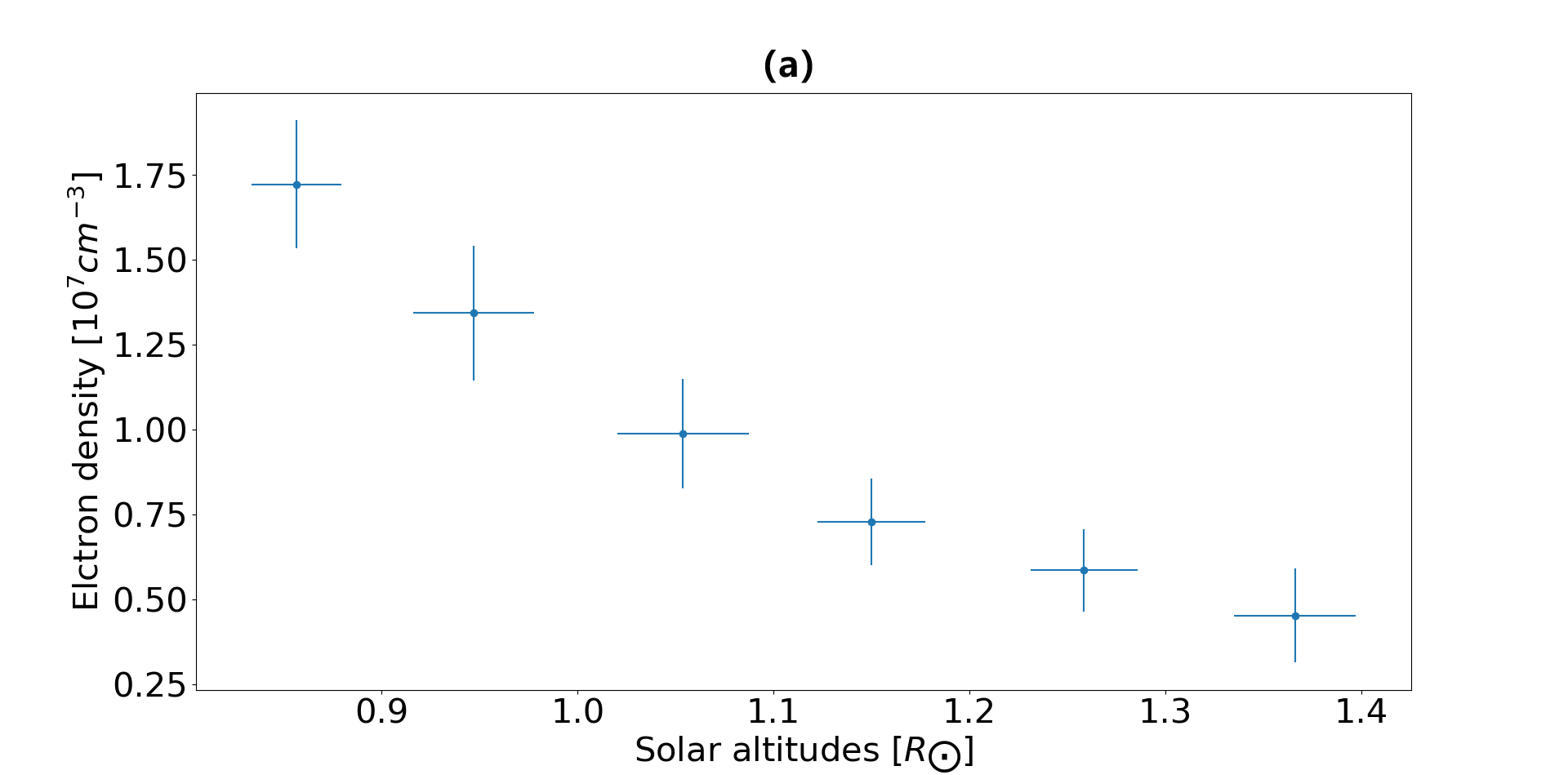}
 \includegraphics[width=0.7\textwidth,trim=8 0 20 10,clip]{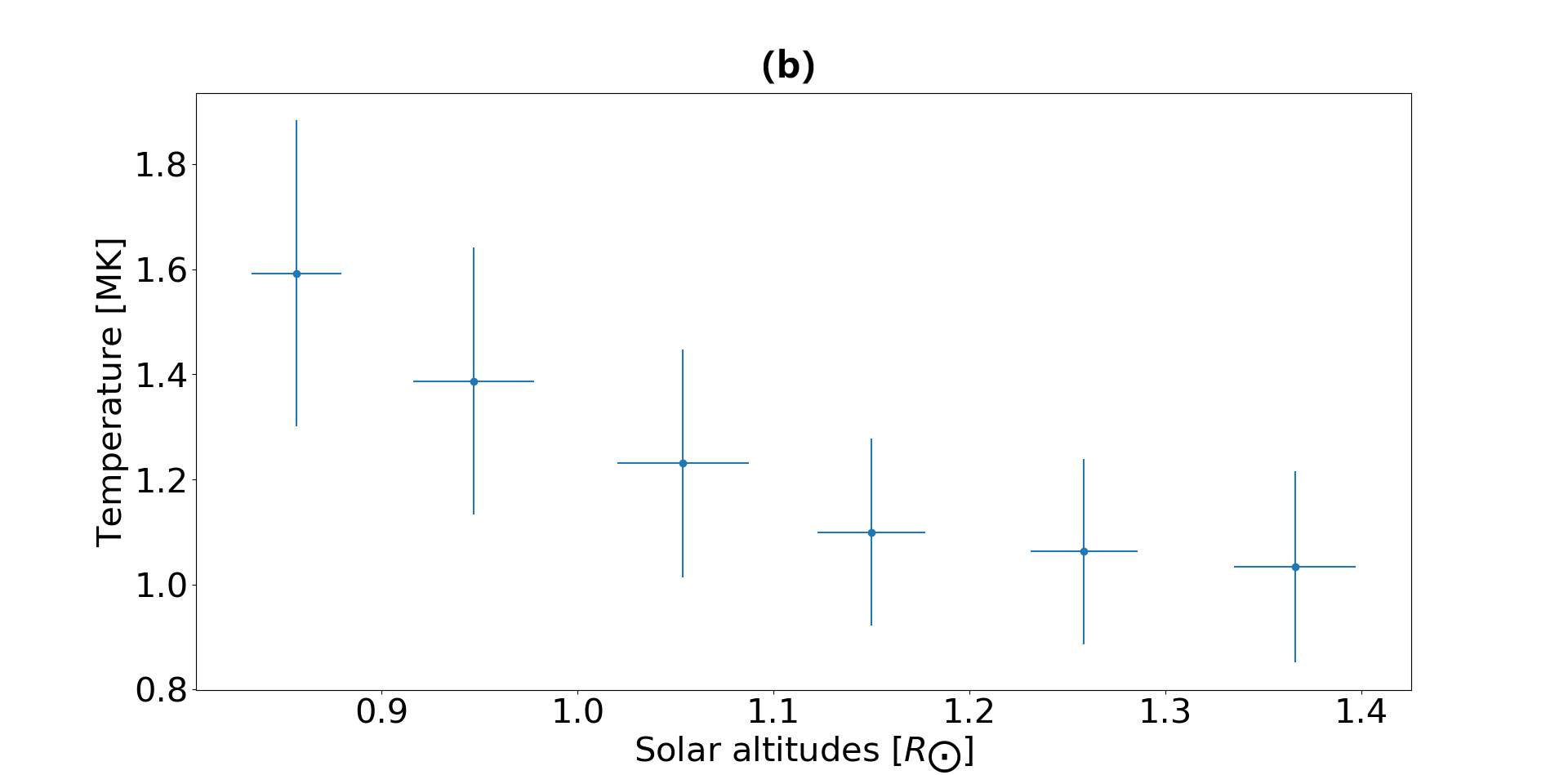}
 \includegraphics[width=0.7\textwidth,trim=8 0 20 10,clip]{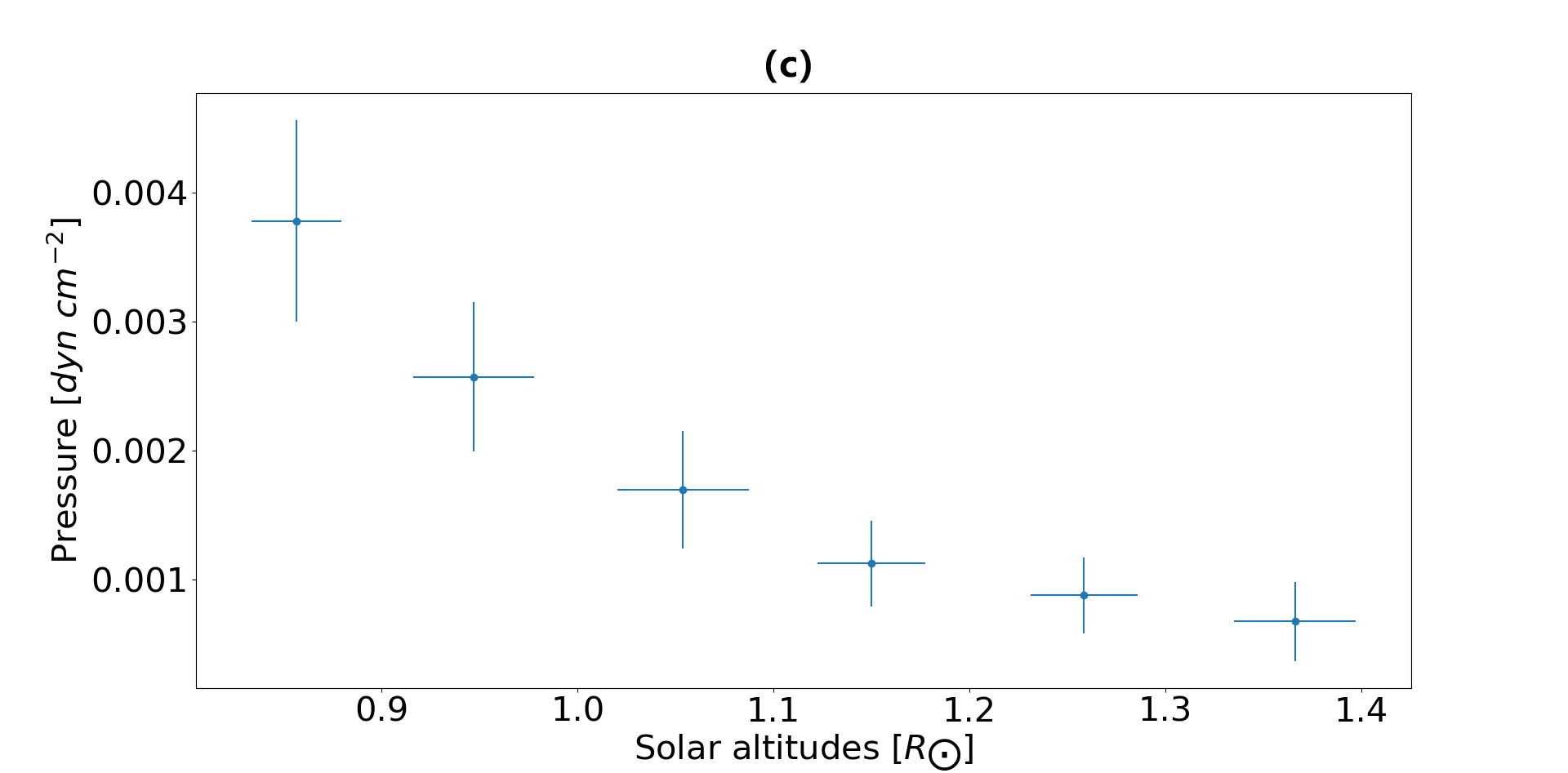}
 \includegraphics[width=0.7\textwidth,trim=8 0 20 10,clip]{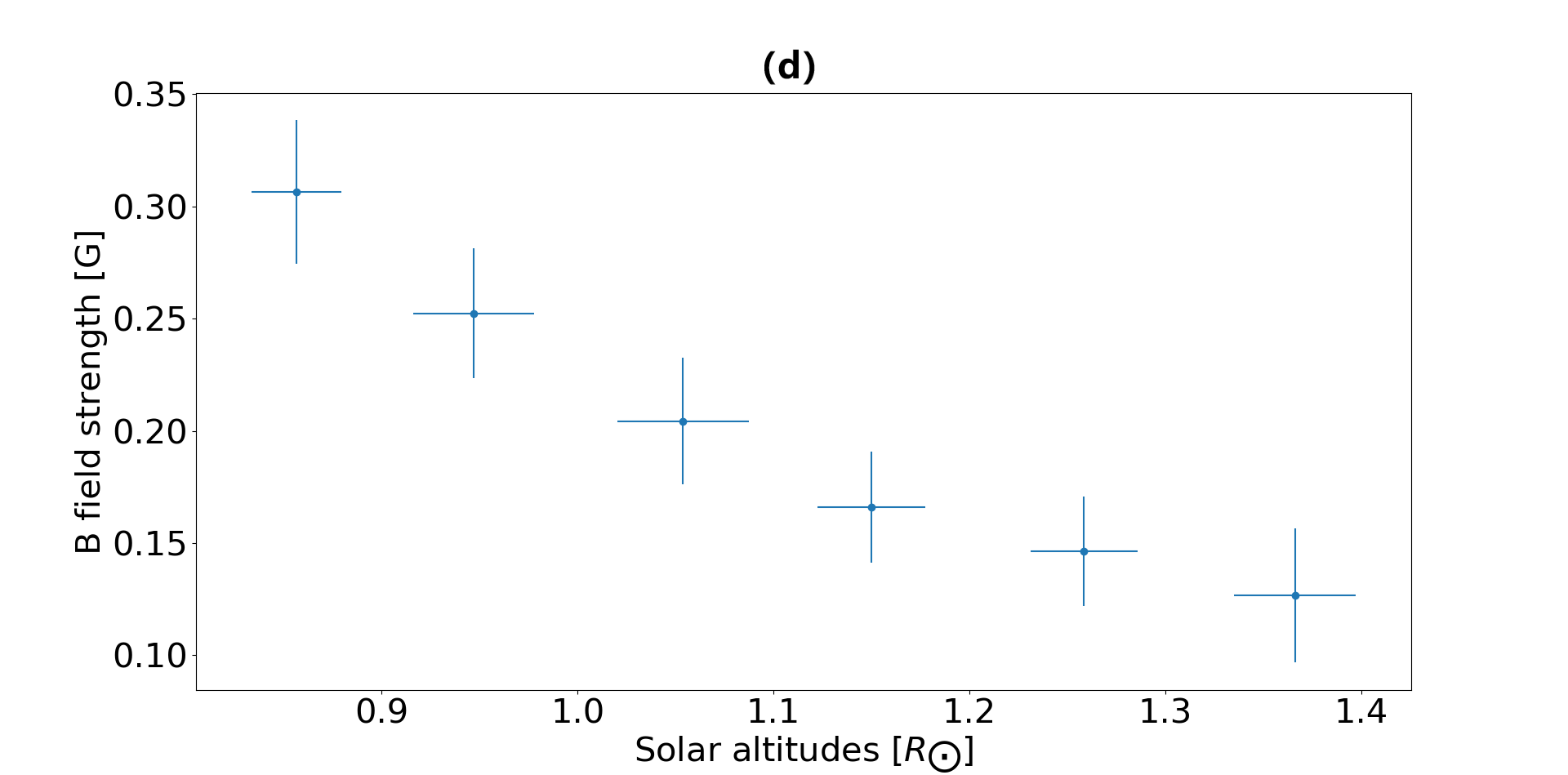}
\caption{Coronal loop plasma parameters inferred from 24 J bursts, averaged every 0.1 solar radius.  \textbf{(a)} electron density, \textbf{(b)} temperature, \textbf{(c)} pressure, \textbf{(d)} minimum magnetic field strength.  Error bars are calculated from the standard deviations to show the spread in results.  The value is the average of each 0.1 solar radius.}
\label{fig: Physical parameter plot} 
\end{figure*}

From all 24 selected J bursts, we found a total of 898 frequency sub-bands ranging between 25 and 80\,MHz.  Using the time of maximum flux for all individual sub-bands, we found the average loop altitudes and hence the average electron density model within the loops.  The minimum and maximum solar altitudes above the photosphere were from 0.79 to 1.85\,$\rm R_{\odot}$, respectively. The average density model is shown in Figure \ref{fig: Physical parameter plot}a.  The points were averaged every 0.1\,$\rm R_{\odot}$ from 0.79 to 1.42\,$\rm R_{\odot}$, which contained 94.2\,\% of all maximum flux points in this range.  There were only a few J bursts that extended to higher altitudes.  The electron number density decreases from $2.0 \times 10^{7}\,\rm cm^{-3}$ to $1.6 \times 10^{6}\,\rm cm^{-3}$.  The error bars in Figure \ref{fig: Physical parameter plot} represent the standard deviation of all points considered.

In Equation \ref{eqn:exponential}, the value of the scale height, $\lambda$, is inferred by fitting an exponential curve to the $n_\mathrm{e}(h)$ profile derived from maximum flux points of each selected J burst. The average background coronal density scale height of 24 inferred coronal loop plasma density models is 0.36\,$\rm R_{\odot}$  ($2.5 \times 10^{10} \pm 4.9 \times 10^{9}\,\rm{cm}$). The largest density scale height we derived among 24 J bursts is 0.5\,$\rm R_{\odot}$ ($3.5 \times 10^{10}\,\rm{cm}$), and the lowest is 0.26\,$\rm{R_{\odot}}$ ($1.8 \times 10^{10}\,\rm{cm}$). The average value of $N_0$ is $5.1 \times 10^{9}\,\rm{cm^{-3}}$, from the lowest $5.1 \times 10^{8}$ to the largest $1.8 \times 10^{10}\,\rm{cm^{-3}}$.

\subsection{Coronal Loop Temperature}\label{sec:Coronal loop temperature}

The density scale height $\lambda$ is proportional to the electron temperature T and inversely proportional to the gravitational acceleration of the Sun.  Assuming the solar coronal plasma is in hydrostatic equilibrium, the density scale height $\lambda$, as determined from the exponential density model in Equation \ref{eqn:exponential}, is given by \citep[see][]{Aschwanden:1992aa,aschwanden2001temperature}

\begin{equation} \label{eqn:densitySH}
\lambda = \frac{1+\alpha}{\beta} \frac{\mathrm{k_B} T}{\mathrm{m_P} g_\odot},
\end{equation}

\noindent
where $\alpha=1.22$ is the ratio of electron to proton number, and $\beta=1.44$ is the mean molecular weight, $\mathrm{k_B}$ is the Boltzmann constant, $\mathrm{m_P}$ is the mass of a proton, and $g_\odot$ is the gravitational acceleration.  In contrast to the U-burst studied at the high frequencies by \citet[e.g.][]{Aschwanden:1992aa}, we analysed J bursts at much lower frequencies, and hence much larger coronal altitudes.  Whilst \citet[e.g.][]{Aschwanden:1992aa} assumed the value of $g_\odot$ at 1\,$\rm{R_\odot}$ from the solar centre, we considered $g_\odot$ as a function of solar altitude such that

\begin{equation} \label{eqn:gsun}
g_\odot (r) = \frac{\mathrm{GM_\odot}}{r^2},
\end{equation}

\noindent
where $\mathrm{G}$ is the gravitational constant, $\mathrm{M_\odot}$ is the mass of the Sun, and $r$ is the distance from the centre of the Sun, such that $r=h+1$. 

For each selected J burst, we determined the coronal loop temperature distributions as a function of solar altitude by combining Equation \ref{eqn:densitySH} and Equation \ref{eqn:gsun} such that

\begin{equation} \label{eqn:Temperature}
T(r) = \frac{\beta}{1+\alpha} \frac{\mathrm{m_P} \lambda}{\mathrm{k_B}}\frac{\mathrm{GM_\odot}}{r^2},
\end{equation}

\noindent
where $\lambda$ is the density scale height value inferred from the selected J burst. Similar to the gravitational acceleration, the loop temperature is inversely proportional to the square of solar altitude. 

By analysing all 24 J bursts, using the 898 frequency sub-bands, the lowest temperature was 0.7\,MK, and the highest temperature was 2.1\,MK. The loop-top temperature decreases from the lower to the upper solar corona. Figure \ref{fig: Physical parameter plot}b shows the average temperature is 1.6 MK at the beginning around 0.86\,$\rm{R_\odot}$ and then drops to 1.0\,MK around 1.37\,$\rm{R_\odot}$ above the photosphere.

\subsection{Coronal Loop Pressure}\label{sec:Coronal loop pressure}

To obtain the loop pressure, $P(h)$, we applied the ideal gas law

\begin{equation} \label{eqn:GASlaw}
P(h) = n_\mathrm{e}(h)  \mathrm{k_b}  T(h),
\end{equation}

\noindent
using our previously deduced background electron density profile $n_\mathrm{e}(h)$ and the temperature profile $T(h)$. Figure \ref{fig: Physical parameter plot}c shows the average loop pressure decreasing with increasing altitude, from $3.8 \times 10^{-3}\,\rm{{dyn \ cm}^{-2}}$ aroud 0.86\,$\rm{R_\odot}$ to $0.7 \times 10^{-3}\,\rm{{dyn \ cm}^{-2}}$ around 1.37\,$\rm{R_\odot}$.

\subsection{Coronal Loop Magnetic Field Strength}\label{sec:Coronal loop magnetic field strength}

It is generally accepted that the magnetic pressure is larger than thermal pressure in the solar corona. This is the reason that accelerated particles in the solar corona propagate along magnetic field lines as they are ``frozen" in magnetic fields. The plasma beta, $\beta_\mathrm{p}$, presents the ratio of the thermal pressure, $P_{\mathrm{thermal}}$, to the magnetic pressure, $P_{\mathrm{magnetic}}$, which is less than one in the solar corona:

\begin{equation} \label{eqn:plasmabeta}
\beta_\mathrm{p} = \frac{P_{\mathrm{thermal}}}{P_{\mathrm{magnetic}}} = \frac{n_\mathrm{e}(h) \ \mathrm{k_b} \ T_\mathrm{e}(h)}{B(h)^2 / 8 \pi} < 1,
\end{equation}
where $B(h)$ is the magnetic field strength.  In a similar way to \citet{Aschwanden:1992aa}, we can estimate the upper bound of the magnetic field strength using
\begin{equation} \label{eqn:magfield}
B(h) > [8\pi n_\mathrm{e}(h) \mathrm{k_b} T_\mathrm{e}(h)]^{0.5}.
\end{equation}
We note that $B(h)$ is proportional to the square root of the plasma pressure and thus has the same trend as the other parameters.  Figure \ref{fig: Physical parameter plot} d shows magnetic field strengths of 0.31\,G around 0.86\,$\rm{R_\odot}$ to 0.13\,G around 1.37\,$\rm{R_\odot}$.

\section{Discussions}\label{sec:Discussions}

\subsection{Type-III and J Bursts Exciter Velocities}\label{sec:type-III and type J bursts exciter velocity}

By analysing 27 type-III and 27 J bursts, we found that the average exciter velocities were comparable (see Table \ref{tab:velocities}).  There is a small increase in the J burst velocities, but any difference between the bursts lies within the standard deviation.  The larger velocities under harmonic emission are related to the decreased gradient of the coronal density model at higher altitudes.  Given the same density model assumed for the type-III bursts and the initial part of the J bursts, the radio drift rates are similar.  This result reinforces the previous hypothesis that both radio bursts are signatures of electron beams propagating along magnetic loops, with the change in J burst drift rate indicating electron propagation near the loop apex. 

That we do not see much change in the exciter velocity suggests there is no discernible difference in the acceleration properties of the corresponding electron beams.  We might have expected a difference because the larger magnetic structures (``open" loops) that are responsible for type-III bursts are likely to have smaller background electron densities and temperatures, in a similar way that coronal holes have cooler, less dense plasma to the surrounding corona.  It may be the case that the particle acceleration in the radio noise storms occurs at the same region and, accelerated beams have access to different magnetic structures as the particle acceleration evolves over time. Whatever the case, models of radio noise storm acceleration need to take into account the similar electron beam velocities from J bursts and type-III bursts.

\subsection{Coronal Loop Physical Parameters}\label{sec:Comparing large and small loops physical parameters}

\begin{center}
\begin{table*} \centering
\caption{Average plasma parameters of large and small coronal loops, near their apex.  Large coronal loop parameters were estimated from our 24 low frequency J bursts, at an altitude of 1.37\,$\rm{R_{\odot}}$.  Small coronal loop parameters were estimated from three high frequency U bursts analysed by \citet{Aschwanden:1992aa}, at altitudes of 0.18\,$\rm{R_{\odot}}$.}
\begin{tabular}{ c c c }

\hline\hline

Plasma Parameters &	Large Coronal Loops    &   Small Coronal Loop     \\ \hline
Density scale height  &	$0.36 \pm 0.07\,\rm{ R_{\odot}}$ 	         &	$0.51 \pm 0.09\,\rm{ R_{\odot}}$\\ 
Density &	$(4.5 \pm 1.4) \ \times \rm{10^{6}\,\rm{cm^{-3}}}$ 	         &	$6.3 \times \rm{10^{9}\,\rm{cm^{-3}}}$\\
Temperature  &	$ 1.0 \pm 0.2$\,MK     &	$7.0 \pm 0.4$\,MK	\\
Pressure  &	 $0.0007 \pm 0.0003\,\rm{dyn\,cm^{-2}}$ 	       &	$6.1 \pm 0.4\,\rm{dyn\,cm^{-2}}$	\\
Minimum magnetic field strength &	 $\ge 0.13 \pm 0.03$\,G	      &	$\ge 12$\,G	\\
 \hline
\end{tabular}
\label{tab:para}
\end{table*}
\end{center}

The average coronal loop altitudes that we found were around 1.37\,$\rm{R_\odot}$.  This altitude is very similar to the 1.45\,--\,1.6\,$\rm{R_\odot}$ loop apex that was inferred from LOFAR imaging observations of two J bursts and one U burst \citep{reid2017imaging}.  The altitudes might be similar as we inferred a background density model (Saito$\times 4.5$) that was similar to these large coronal loops. Nevertheless, the similar loop apex that we found give confidence in our novel technique for estimating the density model in the curved part of these large coronal loops.

The coronal loop altitudes are much smaller than those estimated by \citet{dorovskyy2021solar}.  The main difference between our work and \citet{dorovskyy2021solar} is that they did not infer a density model but assumed an exciter speed of 0.33\,c.  This assumed speed directly led to the large altitudes that they estimated for the coronal loop in the study, and is larger than the speeds that we infer, around 0.2\,c.

The average plasma parameters that we estimated can be found in Table \ref{tab:para}.  We report the average plasma parameters at the loop altitude of 1.37\,$\rm{R_\odot}$, near the apex of most loops, as all parameters vary with altitude.  We also report the average plasma parameters found by \citet{Aschwanden:1992aa} who analysed three U bursts observed by the Very Large Array between 1.3 and 1.7\,GHz, at an approximate altitude of 0.18\,$\rm{R_\odot}$.

For our larger coronal loops, we found similar density scale heights to \citet{Aschwanden:1992aa}.  This result implies that density scale heights are independent on the size of the coronal loop.  However, more coronal loops should be studied, in different active regions and different altitudes (e.g. 0.6\,$\rm{R_\odot}$) before concluding such a result.

The ratio of density scale height to loop altitude in our study is smaller than one, around 0.26.  This is in comparison to being greater than one from the results of \citet{Aschwanden:1992aa}.  As such, we were not able to use the RTV scaling law \citep{rosner1978dynamics} to estimate the plasma temperature using 
\begin{equation} \label{eqn:RTVlaw}
T \approx 1.4 \times 10^{3} (p L)^{-\frac{1}{3}}.
\end{equation}
\citet{Aschwanden:1992aa} compared both methods, finding very similar plasma temperatures, and hence similar plasma pressures.  If indeed the density scale height does not vary significantly with loop height then the RTV law is only useful for more compact coronal loops with altitudes under 0.3\,$\rm{R_\odot}$.

We estimated the loop top temperature of large coronal loops at an average of $1.0 \pm 0.2$\,MK, which is cooler than smaller loops estimated by \citet{Aschwanden:1992aa} at an average of 7\,MK. This agrees with conclusions by \citet{dorovskyy2021solar} that large loops have lower loop top temperature than smaller ones.  The main difference that we infer from \citet{Aschwanden:1992aa} is that we do not take the value of $g_\odot$ to be constant, but instead vary it as a function of solar altitude, $h$.  The temperature dependence is then proportional to $(h+1)^{-2}$ (see also Equation \ref{eqn:Temperature}), and therefore it naturally explains the smaller temperatures at higher altitudes.

The pressure we obtained at the apex of large coronal loops was significantly smaller than found by \citet{Aschwanden:1992aa}.  This occurs naturally from the ideal gas law, primarily based upon the smaller densities, as the difference in plasma temperature is minimal.

We found the minimum magnetic field strength of larger coronal loops to be significantly smaller than was found by \citet{Aschwanden:1992aa}.  Again, this mainly comes directly from the decrease in plasma pressure.  If one assumes a standard dipole approximation of $B(r)\propto r^{-3}$, where $r$ is the distance from the centre of the Sun, the 12\,G that \citet{Aschwanden:1992aa} find at a distance of $r=1.18\,\rm{R_\odot}$ would correspond to 1.48\,G at a distance of $r=2.37\,\rm{R_\odot}$, close to the loop apex we estimate in this study.  This is significantly larger than the 0.13\,G we estimated from our data.  Such a result is expected, given the coronal tendency of the magnetic field to decrease exponentially with altitude \citep{Solanki_2006}, due to the increasing volume the magnetic field can fill as $r$ increases.  For an exponential model of $B=B_0\exp(-\mathrm{b}r)$, we can tentatively use our two data points to find the reference magnetic field of $B_0=1.2$\,kG and the scale factor of $b=3.9$.  However, one must remember that the magnetic field varies significantly between different coronal loops, and our magnetic field estimates are only a lower estimate, based upon $\beta<1$. 

\citet{zaitsev2017constancy} discussed two types of coronal loops, which are i) the tube cross-section does not change much with height in the corona and $\beta<1$; ii) the tube cross-section increases with height and $\beta>1$.  As indicated by \citet{zaitsev2017constancy}, the first loop type is typical for most magnetic loops observed in the corona and is why we used this condition to estimate the minimum magnetic field strength.  With a solar wind dependence of $B\propto r^{-2}$ and $B=3\times10^{-5}$\,G at 1\,AU \citep[e.g.][]{verscharen2019multi}, we would predict a magnetic field strength of 0.25\,G at $r=2.37\,\rm{R_\odot}$.  In the case of the second loop type, the magnetic field strength we estimate would be smaller than 0.13\,G.  This does not compare well to our extrapolated value of the magnetic field strength using data from 1\,AU.

\subsection{Factors Affecting Coronal Loop Physical Parameters}\label{sec:Factors affecting coronal loops physical parameters estimation}

\begin{figure} \center  
\includegraphics[width=0.99\columnwidth]{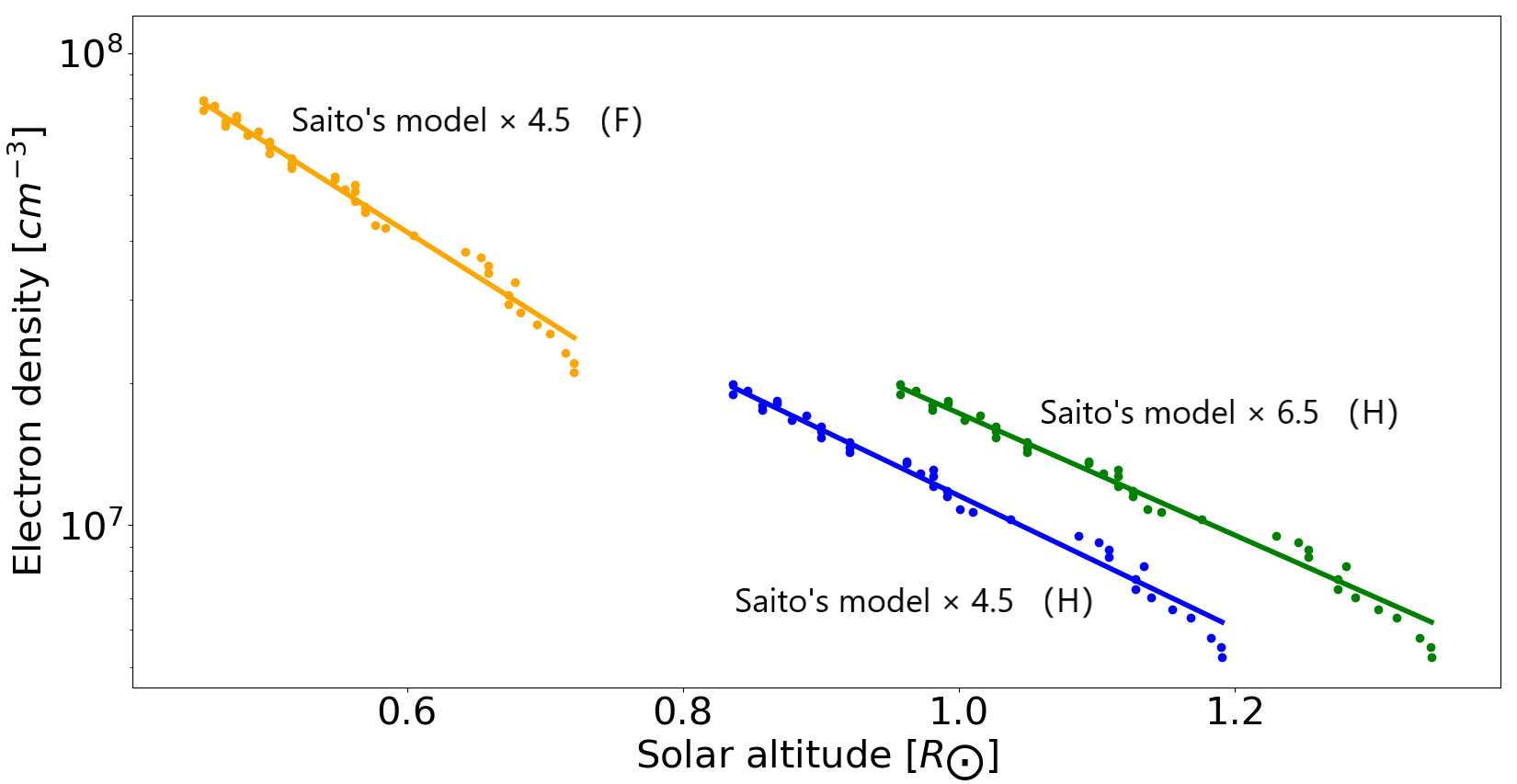}
\caption{Inferred coronal density model from one selected J burst under different assumptions of emission mechanism (Fundamental or Second-harmonic emissions) and reference coronal density model magnitude (Saito's model $\times 4.5$).}
\label{fig:Neh} 
\end{figure}

\subsubsection{The Reference Density Model Magnitude}\label{sec:The reference density model magnitude}

We used the Saito model multiplied by 4.5 for estimating average beam velocities from type-III bursts and the high frequency part of J bursts.  We then used our estimated beam velocities to infer the background electron density models for the curved apex of large coronal loops.  Whilst this modified Saito model was inferred from other J bursts \citep{reid2017imaging}, there is an inherent uncertainty in our results because of this density model assumption.

We can test the effect on our results if we change the multiplication factor that we used on the Saito model.  Figure \ref{fig:Neh} shows how the inferred electron density in the curved apex of a large loop is modified if we used the Saito model multiplied by 6.5.  Because any specific density is now at a larger altitude, the reference altitude of the source position ($l_0$ in Equation \ref{eqn:position}) is higher, and therefore the estimated loop top altitude is higher.  This is evident when comparing the green and blue lines in Figure \ref{fig:Neh}, where we have an increase in apex of the coronal loop from 1.19\,$\rm{R_\odot}$ to 1.34\,$\rm{R_\odot}$ above the photosphere.

Using the Saito model multiplied by 6.5 instead of 4.5 leads to higher beam velocities deduced from the radio bursts.  When using larger magnitude density models, the range of frequencies that we are analysing correspond to densities that are now farther away from the Sun, where the spatial gradient of the density model is smaller.  This leads to longer distances between any given density, and hence the beam needs to travel faster over these long distances as the travel time inferred from the dynamic spectrum is fixed.  The result of these higher beam velocities means that using the Saito model multiplied by 6.5 increases our inferred average density scale heights from 0.35\,$\rm{R_\odot}$ to 0.38\,$\rm{R_\odot}$.  This change in density scale height is not very significant in our altitude range.  With loop temperature, pressure and magnetic field being proportional to the density scale height, the exact density model does not modify significantly the derived plasma parameters at a given altitude.  What has a significant effect on plasma parameters at the loop apex is the change in altitude of the coronal loop, as the temperature is proportional to $r^{-2}$, where $r$ is the distance from the centre of the Sun. 

\subsubsection{Emission Mechanism}\label{sec:Emission mechanism assumptions}

We used the assumption of second-harmonic emission to obtain our results, as second-harmonic emission is more prevalent during type-III solar noise storms in the upper corona (see \citet{kai1985storms} and Section  \ref{sec:introduction}).  Moreover, during the solar radio noise storm we analysed, many J bursts showed ``Fundamental--Harmonic (F--H)" pairs structure on the LOFAR dynamic spectrum.  This structure defined by both fundamental and harmonic components appears together, with the fundamental components at a lower frequency range than the harmonic. In this case, we analysed the harmonic components, because the fundamental components typically showed more structure. However, we cannot be completely certain that all of our radio bursts in this study are generated by harmonic emission.

If we assume fundamental emission for all of our bursts, the corresponding electron densities that the electron beam travels through will be higher.  This means that the altitudes will be lower.  Figure \ref{fig:Neh} shows the derived electron density in the curved apex of a large loop if we assume fundamental emission.  We can clearly see the densities are higher and the altitudes are lower.  This is evident when comparing the yellow and blue lines in Figure \ref{fig:Neh}, where we have an decrease in apex of the coronal loop from 1.19\,$\rm{R_\odot}$ to 0.72\,$\rm{R_\odot}$.

The effect of assuming fundamental emission for all of our J bursts reduces the derived loop apex height, from from 1.37\,$\rm{R_\odot}$ to 0.85\,$\rm{R_\odot}$. The average scale height is reduced from 0.36\,$\rm{R_\odot}$ to 0.27\,$\rm{R_\odot}$. Even with a decrease in the average scale height, the reduced altitude of the loop apex means that the 24 J bursts average loop apex temperature increased from around 1\,MK to 1.3\,MK. The average plasma density at the loop apex is changed from $0.45\times 10^7\,\rm{{cm}^{-3}}$ to $1.8\times 10^7\,\rm{{cm}^{-3}}$. Moreover, under the fundamental emission assumption compared to harmonic, the average loop top pressure dramatically increases from $0.07 \times 10^{-2}\,\rm{dyn\,cm^{-2}}$ to $0.3 \times 10^{-2}\,\rm{dyn\,cm^{-2}}$. Average value of minimum magnetic field strength increases from 0.13\,G to 0.28\,G.  The increase in temperature, density, pressure, and magnetic field are primarily driven by the smaller loops that are inferred by assuming fundamental emission despite the decreases in the average density scale height.

\subsubsection{Loop Top Geometry}\label{sec:Loop top geometry}

\begin{figure} \center  
\includegraphics[width=0.99\columnwidth]{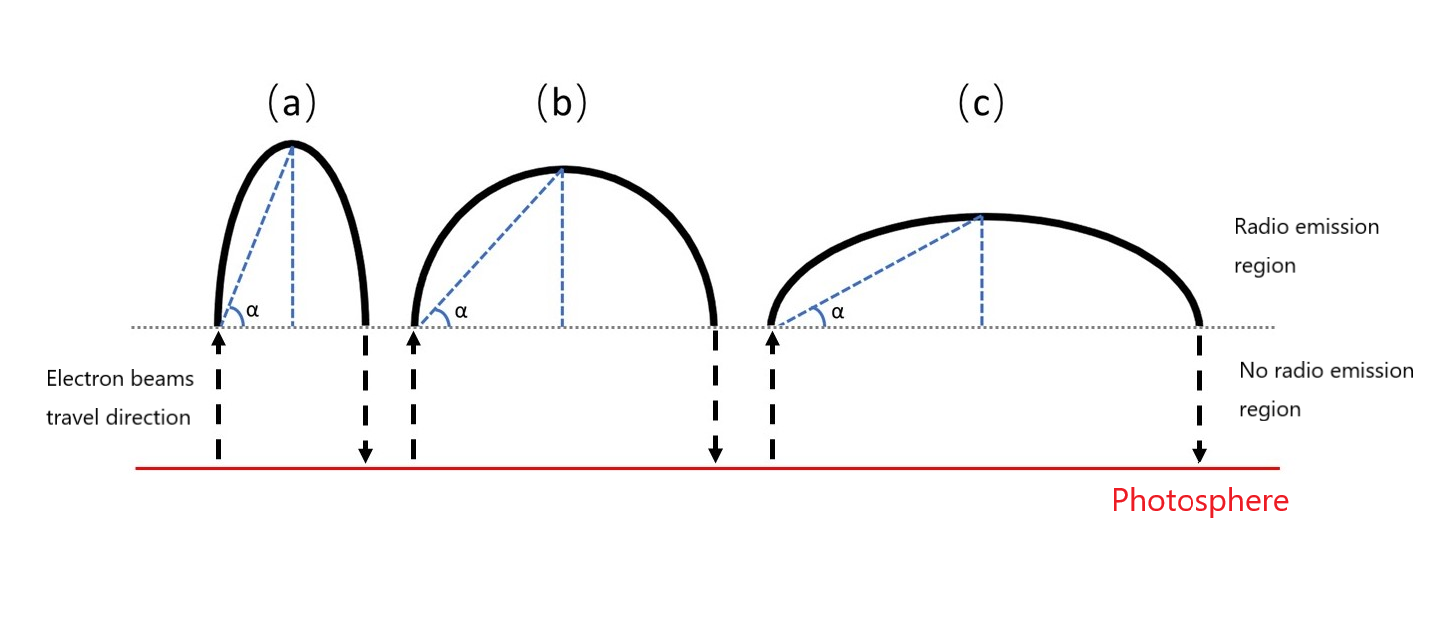}
\caption{Three scenarios of loop top geometry assumptions: \textbf{(a)} A tall semi-ellipse: $\tan(\alpha) > 1$ ; \textbf{(b)} A semi-circle: $\tan(\alpha) = 1$; \textbf{(c)} A flat semi-ellipse: $\tan(\alpha) < 1$.}
\label{fig:loopgeo} 
\end{figure}

In Section \ref{sec:Inferred background density model for the curvature part of the loop}, we introduced the idea that the electron beam travel distance profile along the solar altitude $h(t)$ was converted from the distance profile along the loop by assuming the loop top geometry is a semi-circle. However, it is also common to consider that the loop top geometry is a semi-ellipse \citep[e.g.][]{dorovskyy2021solar}. Figure \ref{fig:loopgeo} shows three scenarios of loop top geometry assumption: (a) tall semi-ellipse, (b) semi-circle, (c) flat semi-ellipse. In the real solar corona, the loop top geometry can be varying between these three scenarios.

In the case that the coronal loop top is a tall semi-ellipse (Figure \ref{fig:loopgeo}a), the electron beam travels to higher altitudes.  As such, the inferred background plasma density model will have a larger density scale height and larger loop apex altitudes. The converse is true if we assume the flat semi-ellipse (Figure \ref{fig:loopgeo}c). 

The loop cross-section is also an important characteristic to be considered. As mentioned in Section \ref{sec:Comparing large and small loops physical parameters}, the loop type (ii) increases  cross-section with height, which will cause the density of an electron beam propagating along the loop to decrease.  This scenario will contribute towards the reduction of radio waves being produced by the electron beam at the top of the loop, and was discussed in depth by \citet{reid2017imaging}. The loop cross-section can be estimated by imaging radio sources. However, exactly what shape coronal loops take near their apex is beyond the scope of this article and will be a subject of a future study.

\section{Conclusions}\label{sec:Conclusions}
This study estimated solar accelerated electron beam velocities by analysing type-III and J bursts, identified from the same solar radio noise storm observed by LOFAR. After comparisons, we found that the kinetic properties of the electron beams are similar while travelling in ``open" or closed flux tubes during the noise storm. This is the first comparison of type-III and J bursts exciter velocities during the same solar radio noise storm.  

Solar J bursts also provide us with an excellent way of estimating the physical parameters of large coronal loops, where the density is too low for typical UV or X-ray diagnostics, and typical coronal density models are not relevant.  We described a novel technique for estimating the background density model of large coronal loops just using the spectroscopic data from a radio dynamic spectrum of a solar J burst.  Using data from 24 J bursts, we inferred the average background density model of coronal loops in an active region.  The coronal loop apexes were situated at an average height of 1.37\,$\rm{R_\odot}$, with an average scale height of 0.36\,$\rm{R_\odot}$.  We then used the scale height to infer average loop plasma temperatures (1\,MK), pressures ($2 \times 10^{-3}\,\rm{dyn\,cm^{-2}}$), and minimum magnetic field strengths (0.22\,G).  These plasma parameters are significantly different from those inferred for smaller coronal loops by \citet{Aschwanden:1992aa} using higher frequency J bursts.  The plasma parameters that we estimated are dependent upon the assumptions that we made about the coronal loop geometry, the background density model and the emission mechanism.  However, we have showed that our derived plasma parameters are not too significantly affected by modifications in these assumptions and should hold as some of the first predictions of the plasma state in such large coronal loops.  Our study has focused on the high resolution spectroscopic data from LOFAR but found similar results for when LOFAR imaging spectroscopy was used \citep{reid2017imaging}. A further study would be interesting to compare our new technique with a density model estimated from interferometric images, taking into account radio scattering effects \citep[e.g.][]{kontar2019anisotropic}, to test the assumption of the electron beam velocity remaining constant as it traversed the apex of the coronal loop.


\begin{acks}
 This article is based on data obtained with the International LOFAR Telescope (ILT). LOFAR (van Haarlem et al. 2013) is the Low Frequency Array designed and constructed by ASTRON. We thank the staff of ASTRON and the LOFAR KSP group. J. Zhang thanks the Mullard Space Science Laboratory, especially the solar group, who helped and encouraged J. Zhang through this project. H. Reid acknowledges funding from the STFC Consolidated Grant ST/W001004/1. V. Krupar acknowledges the support by NASA under grants \texttt{18-2HSWO218\_2-0010} and \texttt{19-HSR-19\_2-0143}. B. Dabrowski and A. Krankowski thank the National Science Centre, Poland for granting “LOFAR observations of the solar corona during PSP perihelion passages” in the Beethoven Classic 3 funding initiative under project number 2018/31/G/ST9/01341. The UWM authors also thank the Ministry of Education and Science (MES), Poland for granting funds for the Polish contribution to the International LOFAR Telescope (agreement  no. 2021/WK/02).
\end{acks}

{\footnotesize
\paragraph*{Disclosure of Potential Conflicts of Interest}
The authors declare that they have no conflicts of interest.
}

{\footnotesize
\paragraph*{Data availability}
The data that support the findings of this study are available from Lofar Long Term Archive, ASTRON, but restrictions apply to the availability of these data, which were used under licence for the current study, and so are not publicly available. Data are however available from the authors upon reasonable request and with permission of ASTRON.
}

\bibliographystyle{spr-mp-sola}
\bibliography{references}

\end{article} 

\end{document}